\documentclass[12pt]{iopart}

\usepackage[dvips]{graphicx}
\usepackage{mathtext}       
\usepackage[english]{babel}
\usepackage{citehack}       

\newcommand{\vt}{\textmd{Ti-6.5Al-2.5Mo-1.5Cr-0.5Fe-0.3Si}}

\newcommand{\lb}{\left(}
\newcommand{\rb}{\right)}
\newcommand{\pd}{\partial}
\newcommand{\half}{\frac{1}{2}}

\begin{document}

\title[Towards analytical solutions of the alloy solidification problem.]
{Towards analytical solutions of the alloy solidification problem.}
\author{E.~N.~Kondrashov\dag
}

\address{\dag JSC VSMPO-AVISMA Corporation, Titanium Alloys Laboratory, Parkovaya Str. 1,
Verkhnyaya Salda, 624760, Sverdlovsk region, Russian Federation}

\ead{evgeniy.kondrashov@vsmpo.ru}

\begin{abstract}

In this paper, an analytical solution of alloy solidification problem is presented.
We develop a special method to obtain an exact analytical solution for mushy zone problem.
The main key of this method is a requirement that thermal diffusivity in the mushy zone
to be constant. From such condition we obtain an ordinary differential equation for
liquid fraction function. Thus the method can be examine as ''a model'' to
achive analytical solution of some unrealistic problems.

An example of solutions is presented: the noneutectic titanium-based alloy solidification.
We provide the comparison of numerical
simulation results with obtained exact solutions. It shown that very simple apparent
capacity-based numerical scheme is provided a good agreement with exact positions
of the solidus and liquidus isoterms, and with temperature profiles also.

Finally, some extensions of the method are outlined.


\end{abstract}





\section{Introduction.}

A general methodolgy of achieving analytical solutions of the
alloy solidification problem is presented in this manuscript.
There is an analytical solution for pure substance (Stefan's problem) and few analytical and
semi-analytical solutions for alloys \cite{alexiades,hu,voller}. We suggest a general methodology
which can provide wide range solutions to test different numerical schemes \cite{alexiades,hu}.

We consider the case when physical properties $\Phi$ (density, heat capacity, or
heat conductivity) in solid and in liquid are constant.
Within mushy zone these properties and the enthalpy depend on temperature as follows
(i.e. obey the lever rule):

\begin{equation}
\Phi (T) = \left[ 1 - \lambda(T) \right] \Phi_{s} + \lambda(T) \Phi_{l},
\end{equation}
where $\Phi_{s}$ and $\Phi_{l}$ are properties in solid and liquid, respectively,
$\lambda(T)$ is volumetric liquid fraction.
Then we rewrite heat transfer equation in the full enthalpy term $H(T)$.
The key idea of the present work is the mushy heat diffusivity requirement to be constant.
\begin{equation}
\alpha(T) = \frac{\kappa(T)}{\frac{dH(T)}{dT}} = \alpha_{sl} = const,
\end{equation}
where $\kappa(T)$ is heat conductivity.
From this condition we can find liquid fraction $\lambda(T)$ by means of which we are able to
linearize an initial energy conservation equation. Thus, the following methodology is
\begin{enumerate}
\item
	To rewrite of the heat equation in the full enthalpy term.
\item
	To require of the thermal diffusivity to be constant in the solid, mushy and liquid zone.
\item
	Condition $\alpha(T) = \alpha_{sl} = const$ is ordinary differential equation for liquid fraction
	$\lambda = \lambda(T)$. Additionaly we require $\lambda(T_{l}) = 1$.
\item
	To solve this ODE and find $\lambda= \lambda(T, \alpha_{sl})$.
\item
	To impose additional condition $\lambda(T_{s}, \alpha_{sl}) = \lambda_{0}$.
	For $\lambda_{0} = 0$ we have
	noneutectic alloy, and for $\lambda_{0} \ne 0$ -- eutectic. From this condition
	we find $\alpha_{sl}$.
\item
	Now we have the heat equation with constant-peace coefficients and we can it solve easy.
\end{enumerate}
It needs to note, that this problem cannot be solved with {\em well defined} (predefined)
$\lambda(T)$ function, instead of the function $\lambda = \lambda(T)$ is determined from
linearisation conditions.

\section{The linearisation of the heat equation.}

We will solve an energy conservation equation

\begin{eqnarray}
\frac{\pd H}{\pd t} = div \lb k(T) grad \, T \rb,
\label{h-eq}
\end{eqnarray}
where full enthalpy $H$ is

\begin{equation}
\fl H(T) = \rho(T) \left[1 - \lambda(T)\right] \int\limits_{0}^{T} C_{s}(\zeta)d\zeta +
\rho(T)\lambda(T)\int\limits_{0}^{T}C_{l}(\zeta)d\zeta +
\rho(T) \lambda(T) L,
\label{enthalpy-definition}
\end{equation}
where $C_{s}$, $C_{l}$ is specific heat in solid and in liquid,
$L$ is latent heat of fusion, $\rho=\rho_{s}=\rho_{l}$ is density, which all are
constants.
We express the heat conductivity in the ''mixture'' form

\begin{equation}
\kappa(T) = \left[ 1 - \lambda(T) \right] \kappa_{s} + \lambda(T) \kappa_{l},
\label{cond_def}
\end{equation}
where $\kappa_{s}$ and $\kappa_{l}$ are constant heat conductivity in solid and liquid,
respectively.

\noindent Taking into account the expression
\begin{equation}
grad \, T = \frac{grad \, H}{\frac{dH}{dT}}
\end{equation}
the Eq.~(\ref{h-eq}) can be rewritten in the general form

\begin{equation}
\frac{\pd H}{\pd t} = div \lb \alpha(T) grad \, H \rb,
\label{h-eq-new}
\end{equation}
where $\alpha(H)$ is thermal diffusivity, which defined as
\begin{equation}
\alpha(T) = \frac{\kappa(T)}{\frac{dH(T)}{dT}},
\end{equation}

\noindent If $\kappa(T)$ and $dH(T)/dT$ depend on temperature arbitrary manner then
Equation~(\ref{h-eq-new}) is nonlinear. To achieve an analitical solution we need to require the
thermal diffusivity to be constant in all regions (solid, mushy and liquid).

\begin{equation}
\label{a}
\alpha(T) = \cases{
\alpha_{s} = const & for $T < T_{s}$ \\
\alpha_{sl} = const & for $T_{s} \leq T \leq T_{l}$ \\
\alpha_{l} = const & for $T > T_{l}$\\}
\end{equation}
In our case $\alpha_{s} = \kappa_{s}\rho C_{s}$ and $\alpha_{l} = \kappa_{l}/\rho C_{l}$
are constant by definition.
For the derivation of mushy enthalpy (the apparent capacity $\times$ density) we get:

\begin{equation}
\frac{1}{\rho}\frac{dH(T)}{dT} =
\left[ 1-\lambda \right] C_{s} +
\lambda C_{l} +
\left[ (C_{l} - C_{s})T + L \right] \frac{d\lambda(T)}{dT}.
\end{equation}

then from the mushy part of Eq.~(\ref{a}) we obtain an ordinal differential equation for
$\lambda(T)$

\begin{equation}
\left[ 1 + pT \right] \frac{d\lambda(T)}{dT} + a \lambda(T) + b  = 0,
\label{ODE_g}
\end{equation}
where we denote

\begin{eqnarray}
a = \frac{ a_{sl} \rho (C_{l} - C_{s}) - (\kappa_{l} - \kappa_{s}) }
{a_{sl}\rho L}, \\
b = \frac{ a_{sl} \rho C_{s} - \kappa_{s} }
{a_{sl} \rho L}, \\
p = \frac{C_{l} - C_{s}}{L}.
\end{eqnarray}
We require

\begin{equation}
\lambda(T_{l}) = 1,
\label{ODE_g_bc}
\end{equation}
where $T_{l}$ is a liquidus temperature. Solution of Eqs. (\ref{ODE_g}) and (\ref{ODE_g_bc}) is

\begin{equation}
\lambda(T) = - \frac{b}{a} + \frac{a+b}{a} \lb \frac{1+pT_{l}}{1+pT} \rb^{\frac{a}{p}}.
\end{equation}

\noindent It needs to determine an additional condition for $\lambda(T)$ function, namely
to define the liquid fraction value at solidus temperature

\begin{equation}
\label{g0}
\lambda(T_{s}) = \cases{
0 & for noneutectic alloy \\
\lambda_{0} & for eutectic alloy\\}
\end{equation}

\noindent To obtain the analytical solution of Eq.~(\ref{h-eq-new}) 
we need to solve Eq.(\ref{g0}) to find root $a_{sl}$.
Then we need to solve Eq.~(\ref{h-eq-new}) with suitable initial and boundary conditions.

The enthalpy of the system (\ref{enthalpy-definition}) we may
design

\begin{equation}
\label{enthalpy}
\fl \frac{H(T)}{\rho} = \cases{
C_{s} T & for $T < T_{s}$ \\
C_{s}T + \lambda(T)\left[ (C_{l} - C_{s})T + L \right] & for $T_{s} \leq T \leq T_{l}$ \\
C_{l}T + L & for $T > T_{l}$\\}.
\end{equation}

In future we need the value $dT/dH$:

\begin{equation}
\label{dTdH}
\fl \rho \frac{dT}{dH} = \cases{
\frac{1}{C_{s}} & for $T < T_{s}$ \\
\frac{1}{ \left[1-\lambda \right]C_{s} + \lambda C_{l} + \left[ (C_{l}-C_{s}T+L)
\right]\frac{d\lambda}{dT} }
& for $T_{s} \leq T \leq T_{l}$ \\
\frac{1}{C_{l}} & for $T > T_{l}$\\}.
\end{equation}

It should be note that some expressions with $d\lambda/dT$ may be written in more simplified
form versus $\lambda$, for example:

\begin{equation}
\frac{d\lambda}{dT} = - \frac{a\lambda+b}{1+pT} =
- \frac{a\lambda+b}{(C_{l}-C_{s})T+L}L,
\end{equation}
and the combination

\begin{equation}
\left[ (C_{l}-C_{s})T + L \right]\frac{d\lambda}{dT} = - \lb a\lambda + b \rb L.
\end{equation}

\section{An analytical solution for enthalpy.}

\noindent We will examine the simple problem

\begin{eqnarray}
\frac{\pd H}{\pd t} = \frac{\pd}{\pd x} \lb \alpha(H) \frac{\pd H}{\pd x} \rb, \\
\label{main-eq}
H(t = 0) = H(T_{init}), \\
\left. H \right|_{x=0} = H_{out} = H(T_{out}), \\
\left. H \right|_{x=\infty} = H_{init} = H(T_{init}).
\end{eqnarray}

\noindent The solution of these equations with constant-piece function $\alpha(H)$
can be easy find \cite{crank}. To solve this equation we divide whole region $[0,\infty)$
into three subintervals $[0,X_{s})$, $[X_{s},X_{l}]$ and $(X_{l}, \infty)$ ($X_{s}$ and
$X_{l}$ are solidus and liquidus positions, respectively).
Moreover, we assume that (the similarity solution):

\begin{equation}
X_{s}(t) = k_{s} \sqrt{t}, \qquad X_{l}(t) = k_{l} \sqrt{t},
\label{Xs-Xl-tau}
\end{equation}
where $k_{s}$ and $k_{l}$ are constants. Solutions on the subintervals are:

\begin{equation}
\fl H(x,t) = H_{out} + (H_{s} - H_{out})
\frac{erf\left( \frac{x}{2\sqrt{\alpha_{s} t}} \right)}{erf \lb \frac{k_{s}}{2\sqrt{\alpha_{s}}} \rb },
\qquad
x \in [0, X_{s}),
\label{h_solution_s}
\end{equation}

\begin{equation}
\fl H(x,t) = \frac{ (H_{l}-H_{s}) erf \lb \frac{x}{2\sqrt{\alpha_{sl} t}} \rb
+ H_{s} erf \lb \frac{k_{l}}{2\sqrt{\alpha_{sl}}} \rb -
H_{l} erf \lb \frac{k_{s}}{2\sqrt{\alpha_{sl}}} \rb     }
{  erf \lb \frac{k_{l}}{2\sqrt{\alpha_{sl}}} \rb  -
erf \lb \frac{k_{s}}{2\sqrt{\alpha_{sl}}} \rb  },
\quad
x \in [X_{s}, X_{l}],
\label{h_solution_sl}
\end{equation}

\begin{equation}
\fl H(x,t) = 
H_{init} - (H_{init} - H_{l}) 
\frac{  erfc \lb \frac{x}{2\sqrt{\alpha_{l}t}} \rb  }
{  erfc \lb \frac{k_{l}}{2\sqrt{\alpha_{l}}} \rb   },
\quad
x \in (X_{l}, \infty),
\label{h_solution_l}
\end{equation}
where we defined

\begin{eqnarray}
H_{s} = H(T_{s}) = \rho C_{s} T_{s},\\
H_{l} = H(T_{l}) = \rho ( C_{l} T_{l} + L ).
\end{eqnarray}

\noindent By using the two conditions at the interfaces (the first one from which
is Stefan's condition at the solidus (eutectic) point):

\begin{equation}
\left. \alpha_{s} \frac{\pd H}{\pd x} \right|_{x=X_{s}-0} =
\left. \alpha_{sl} \frac{\pd H}{\pd x} \right|_{x=X_{s}+0}
+
\rho \lambda_{0} L \frac{dX_{s}(t)}{dt},
\label{cond-s}
\end{equation}

\begin{equation}
\left. \alpha_{sl} \frac{\pd H}{\pd x} \right|_{x=X_{l}-0} =
\left. \alpha_{l} \frac{\pd H}{\pd x}
\right|_{x=X_{l}+0},
\label{cond-l}
\end{equation}
we derive the following two equations from which to evaluate $k_{s}$ and $k_{l}$:

\begin{equation}
\fl \frac{ \sqrt{\alpha_{s}}(H_{s} - H_{out}) exp\lb -\frac{k_{s}^{2}}{4\alpha_{s}} \rb }
{erf\lb \frac{k_{s}}{2\sqrt{\alpha_{s}}} \rb}
-
\frac{ \sqrt{\alpha_{sl}} (H_{l} - H_{s}) exp\lb - \frac{k_{s}^{2}}{4\alpha_{sl}} \rb }
{erf\lb \frac{k_{l}}{2\sqrt{\alpha_{sl}}} \rb - erf\lb \frac{k_{s}}{2\sqrt{\alpha_{sl}}} \rb }
= \frac{\sqrt{\pi}}{2} \rho \lambda_{0} L k_{s},
\label{cond-s-real}
\end{equation}

\begin{equation}
\fl \frac{ \sqrt{\alpha_{sl}} (H_{l} - H_{s}) exp\lb - \frac{k_{l}^{2}}{4\alpha_{sl}} \rb }
{erf\lb \frac{k_{l}}{2\sqrt{\alpha_{sl}}} \rb - erf\lb \frac{k_{s}}{2\sqrt{\alpha_{sl}}} \rb }
-
\frac{ \sqrt{\alpha_{l}}(H_{init} - H_{l}) exp\lb -\frac{k_{l}^{2}}{4\alpha_{l}} \rb }
{erfc\lb \frac{k_{l}}{2\sqrt{\alpha_{l}}} \rb}
= 0.
\label{cond-l-real}
\end{equation}

\section{What we can get from the exact solution? \label{section_temperatures}}

Usualy we have numerical scheme which gives us the temperature, but not enthalpy.
Below we write down formulas for temperature evaluation versus enthalpy and some other
parameters.

\subsection{Solidus and liquidus velocities. \label{section_velocity}}

From Eq.~(\ref{Xs-Xl-tau}) we get the front velocities
\begin{equation}
v_{l}(t) = \frac{d X_{l}(t)}{d t} = \frac{k_{l}}{2 \sqrt{t}},
\qquad
v_{s}(t) = \frac{d X_{s}(t)}{d t} = \frac{k_{s}}{2 \sqrt{t}}.
\label{sl_vels}
\end{equation}

\subsection{Temperature curves. \label{section_curves}}

From Eq.~(\ref{enthalpy}) and Eqs.~(\ref{h_solution_s}) - (\ref{h_solution_l}) we can easy to find:

In the solid ($x < k_{s}\sqrt{t}$):

\begin{equation}
\fl T(x,t) = \frac{1}{\rho C_{s}} \left[ H_{out} +
(H_{s} - H_{out}) \frac{erf\lb \frac{x}{2\sqrt{\alpha_{s}t}} \rb}
{erf\lb \frac{k_{s}}{2\sqrt{\alpha_{s}}} \rb} \right].
\end{equation}

In the mushy zone ($k_{s}\sqrt{t} \leq x \leq k_{l}\sqrt{t}$) we need to solve
nonlinear equation to get $T=T(x,t)$:

\begin{eqnarray}
\fl C_{s}T + \lambda(T) \left[ (C_{l}-C_{s})T + L \right] \nonumber \\ 
= 
\frac{1}{\rho}
\frac{ (H_{l}-H_{s}) erf \lb \frac{x}{2\sqrt{\alpha_{sl}t}} \rb
+ H_{s} erf \lb \frac{k_{l}}{2\sqrt{\alpha_{sl}}} \rb -
H_{l} erf \lb \frac{k_{s}}{2\sqrt{\alpha_{sl}}} \rb     }
{  erf \lb \frac{k_{l}}{2\sqrt{\alpha_{sl}}} \rb  - erf \lb \frac{k_{s}}{2\sqrt{\alpha_{sl}}} \rb  }.
\end{eqnarray}

In the liquid ($x > k_{l}\sqrt{t}$):

\begin{equation}
\fl T(x,t) = \frac{1}{\rho C_{l}} \left[ H_{init} - (H_{init} - H_{l}) 
\frac{  erfc \lb \frac{x}{2\sqrt{\alpha_{l}t}} \rb  }
{  erfc \lb \frac{k_{l}}{2\sqrt{\alpha_{l}}} \rb   } \right]
- \frac{L}{C_{l}}.
\end{equation}

\subsection{Temperature gradients. \label{section_gradients}}

From equation

$$
\frac{\pd T}{\pd x} = \frac{d T}{d H} \frac{\pd H}{\pd x}
$$
end from Eq.~(\ref{dTdH}) easy to achive the expressions for temperature gradients:

\begin{equation}
\label{t_grad}
\fl \frac{\pd T}{\pd x} = \cases{
\frac{1}{\rho C_{s}} \frac{\pd H}{\pd x} & $T < T_{s}$ \\
\frac{1}{\rho} \frac{1}{ \left[1-\lambda \right]C_{s} + \lambda C_{l} + \left[ (C_{l}-C_{s})T + L \right]
\frac{d\lambda}{dT} }
\frac{\pd H}{\pd x} & $T_{s} \leq T \leq T_{l}$ \\
\frac{1}{\rho C_{l}} \frac{\pd H}{\pd x} & $T > T_{l}$\\}
\label{t_gradients}
\end{equation}
For enthalpy gradients from Eqs.~(\ref{h_solution_s}) - (\ref{h_solution_l}) we get

\begin{equation}
\label{h_grad}
\fl \frac{\pd H}{\pd x} = \cases{
\frac{H_{s}-H_{out}}{erf\lb \frac{k_{s}}{2\sqrt{\alpha_{s}}} \rb}
\cdot
\frac{e^{-\frac{x^{2}}{4\alpha_{s}t}}}{\sqrt{\pi \alpha_{s} t}} & $H < H_{s}$ \\
\frac{H_{l}-H_{s}}{ erf\lb \frac{k_{l}}{2\sqrt{\alpha_{sl}}} \rb -
erf\lb \frac{k_{s}}{2\sqrt{\alpha_{sl}}} \rb   } \cdot
\frac{e^{-\frac{x^{2}}{4\alpha_{sl}t}}}{\sqrt{\pi \alpha_{sl} t}} & $H_{s} \leq H \leq H_{l}$ \\
\frac{H_{init}-H_{l}}{erfc\lb \frac{k_{l}}{2\sqrt{\alpha_{l}}} \rb} \cdot
\frac{e^{-\frac{x^{2}}{4\alpha_{l} t}}}{\sqrt{\pi \alpha_{l} t}} & $H > H_{l}$\\}
\label{h_gradients}
\end{equation}

Thus finally we have:

\begin{equation}
\label{t_grad_last}
\fl \frac{\pd T}{\pd x} = \cases{
\frac{1}{\rho C_{s}} \cdot \frac{H_{s}-H_{out}}{erf\lb \frac{k_{s}}{2\sqrt{\alpha_{s}}} \rb} \cdot
\frac{e^{-\frac{x^{2}}{4\alpha_{s} t}}}{\sqrt{\pi\alpha_{s} t}} & $x < k_{s}\sqrt{t}$ \\
\frac{ (H_{l}-H_{s})/\left( erf\lb \frac{k_{l}}{2\sqrt{\alpha_{sl}}} \rb - erf\lb 
\frac{k_{s}}{2\sqrt{\alpha_{sl}}} \rb \right) }
{\rho \left[ (1-\lambda)C_{s} + \lambda C_{l} + \left[ (C_{l}-C_{s})T + L \right] \frac{d\lambda}{dT} \right]}
\cdot
\frac{e^{-\frac{x^{2}}{4\alpha_{sl} t}}}{\sqrt{\pi\alpha_{sl} t}}
& $k_{s}\sqrt{t} \leq x \leq k_{l}\sqrt{t}$ \\
\frac{1}{\rho C_{l}} \cdot \frac{H_{init}-H_{l}}{erfc\lb \frac{k_{l}}{2\sqrt{\alpha_{l}}} \rb}
\cdot \frac{e^{-\frac{x^{2}}{4\alpha_{l} t}}}{\sqrt{\pi\alpha_{l} t}}
& $x > k_{l}\sqrt{t}$\\}
\label{t_gradients_last}
\end{equation}

There is a very interesting parameters as temperature gradient in liquid phase at the liquidus
point $G_{l}$. For it we can write down

\begin{equation}
G_{l}(t) = \left. \frac{\pd T(x,t)}{\pd t} \right|_{x=X_{l}+0} =
\frac{1}{\rho C_{l}} \cdot \frac{H_{init}-H_{l}}{erfc\lb \frac{k_{l}}{2\sqrt{\alpha_{l}}} \rb}
\cdot \frac{e^{-\frac{k_{l}^{2}}{4\alpha_{l}}}}{\sqrt{\pi\alpha_{l} t}}.
\label{G_l}
\end{equation}
This parameter controls the type of solidification microstructure.

\subsection{Cooling rate. \label{section_cooling}}

As we can see $G_{l}(t) \sim 1/\sqrt{t}$. Liquidus velocity (\ref{sl_vels}) varies
with time alse as $v_{l}(t) \sim 1/\sqrt{t}$, then cooling rate given by
$G_{l}v_{l} \sim 1/t$. It is easy to show that is so. The cooling rate is defined as

\begin{equation}
\dot{T}(x,t) = \frac{\pd T(x,t)}{\pd t} =
\frac{dT}{dH} \frac{\pd H}{\pd t}
\end{equation}

\begin{equation}
\label{colling_rates}
\fl \frac{\pd T}{\pd t} = \cases{
- \frac{1}{\rho C_{s}} \cdot \frac{H_{s}-H_{out}}{erf\lb \frac{k_{s}}{2\sqrt{\alpha_{s}}} \rb} \cdot
\frac{\alpha_{s} x e^{-\frac{x^{2}}{4\alpha_{s}t}}}{2\sqrt{\pi}(\alpha_{s} t)^{3/2}}
& $x < k_{s}\sqrt{t}$ \\
- \frac{ (H_{l}-H_{s})/\left( erf\lb \frac{k_{l}}{2\sqrt{\alpha_{sl}}} \rb - erf\lb 
\frac{k_{s}}{2\sqrt{\alpha_{sl}}} \rb \right) }
{\rho \left[   (1-\lambda)C_{s} + \lambda C_{l} + \left[ (C_{l}-C_{s})T + L \right] \frac{d\lambda}{dT}  \right]}
\cdot
\frac{\alpha_{sl} x e^{-\frac{x^{2}}{4\alpha_{sl} t}}}{2\sqrt{\pi} (\alpha_{sl} t)^{3/2}}
& $k_{s}\sqrt{t} \leq x \leq k_{l}\sqrt{t}$ \\
- \frac{1}{\rho C_{l}} \cdot \frac{H_{init}-H_{l}}{erfc\lb \frac{k_{l}}{2\sqrt{\alpha_{l}}} \rb}
\cdot \frac{\alpha_{l} x e^{-\frac{x^{2}}{4\alpha_{l} t}}}{2\sqrt{\pi} (\alpha_{l} t)^{3/2}}
& $x > k_{l}\sqrt{t}$\\}
\label{cooling}
\end{equation}

Thus, cooling rate at the liquidus point is given by

\begin{equation}
\dot{T}_{l} = \left. \dot{T} \right|_{x=X_{l}+0} =
- \frac{1}{\rho C_{l}} \cdot \frac{H_{init}-H_{l}}{erfc\lb \frac{k_{l}}{2\sqrt{\alpha_{l}}} \rb}
\cdot \frac{k_{l}e^{-\frac{k_{l}^{2}}{4\alpha_{l}}}}{2 \sqrt{\pi\alpha_{l}} t}
\sim \frac{1}{t}.
\label{dot_T_l}
\end{equation}

The value $\dot{T}_{l}$ is very important, because it defines secondary arm dendrite spacing
\cite{kurz-fischer}.
The another important expression is $G_{l}^{-1/2} v_{l}^{-1/4}$, which defines
primary arm dendrite spacing \cite{kurz}. From Eqs.~(\ref{sl_vels}) and (\ref{G_l}) we can
show that primary arm spacing varies versus time like $\sim t^{3/8}$. However, as it's very
known, after some critical gradient and velocity at the liquidus point will be take a place
columnar to eqiaxed transition \cite{martorano}.

\subsection{Local solidification time \label{section_lst}}

For directional solidificcation, the local solidification time $t_{ls}$ can be estimated from
the following equation:

\begin{equation}
t_{ls}(x) = \left[ \frac{1}{k_{s}^{2}} - \frac{1}{k_{l}^{2}} \right] x^{2},
\end{equation}
where quadratic increasing with $x$ of $t_{ls}$ we have, because the mushy zone lehgth increases
versus $x$ and solidus/liquidus velocities decrease. The local solidification time controls
the some segragation processes in the mushy zone.

\section{Numerical scheme. \label{numerical_scheme}}

We will solve heat transfer equation

\begin{eqnarray}
\rho C(T) \frac{\pd T}{\pd t} =
\frac{\pd}{\pd x} \lb \kappa(T) \frac{\pd T}{\pd x} \rb
\label{main_eq_t}
\end{eqnarray}
which concerns with Eq.~(\ref{main-eq}). Here $\kappa(T)$ edfined by Eq.~(\ref{cond_def})
and mushy heat capacity (so-called {\it apparent capacity}):

\begin{equation}
C(T) = \left[ 1 - \lambda(T) \right] C_{s} + \lambda(T) C_{l} +
\left[ (C_{l}-C_{s})T + L \right] \frac{d\lambda(T)}{dT}.
\end{equation}

The first, we draw the grid with spatial step $h = x_{i+1} - x_{i}$, where $i = 0, 1, ... , N$.
The second, integrating the Eq.~(\ref{main_eq_t}) over the $x \in [(i-\frac{1}{2})h, (i+\half)h]$
we can write heat balance equation as follows

\begin{equation}
\fl \rho C(T_{i}) \frac{\pd T_{i}}{\pd t} h =
\left. \lb \kappa(T)\frac{\pd T}{\pd x} \rb \right|_{x=(i-\half)h}^{x=(i+\half)h}
\approx
\kappa_{i+\half} \frac{T_{i+1}-T_{i}}{h} - \kappa_{i-\half} \frac{T_{i} - T_{i-1}}{h}.
\end{equation}

The left part of this equation we express as

$$
\rho C(T_{i}) \frac{\pd T_{i}}{\pd t}
\approx
\rho C(T_{i}^{n}) \frac{T_{i}^{n+1}-T_{i}^{n}}{\tau},
$$
where $n$ is time index, $\tau$ is time step. Thus the Eq.~(\ref{main_eq_t}) we can write
down in the discrete form as

\begin{equation}
\fl \left[ \frac{\tau\kappa_{i-\half}}{h^{2}} \right] T_{i-1}^{n+1} -
\left[ \frac{\tau}{h^{2}} \lb \kappa_{i-\half}+\kappa_{i+\half} \rb + \rho C_{i} \right]
T_{i}^{n+1} +
\left[ \frac{\tau\kappa_{i+\half}}{h^{2}} \right] T_{i+1}^{n+1} =
- \rho C_{i} T_{i}^{n},
\label{disctrete_eq_t}
\end{equation}
where \cite{patankar}

$$
\kappa_{i-\half} = \frac{2\kappa_{i-1}\kappa_{i}}{\kappa_{i-1}+\kappa_{i}},
\qquad
\kappa_{i+\half} = \frac{2\kappa_{i}\kappa_{i+1}}{\kappa_{i}+\kappa_{i+1}}.
$$

We note, that $C_{i}$ etc are calculated at time $t_{n}$, i.e.

$$
C_{i} = C(T_{i}^{n}),
\qquad
\kappa_{i-1} = \kappa(T_{i-1}^{n})
\qquad
etc.
$$
Boundary conditions are

\begin{equation}
T_{0} = T_{out},
\qquad
T_{N} = T_{init}.
\end{equation}
Because the Eq.~(\ref{disctrete_eq_t}) is tri-diagonal linear system, then its solving is
trivial and we do not discuss this issue here.

\section{Binary alloy solidification: numerical treatment. \label{simple_alloy}}

In this section we consider the solidification of noneutectic titanium-based alloy,
which we can treat as pseudo-binary alloy.
Physical properties of titanium alloy VT3-1 (\vt) are present in the
Table~\ref{table-vt3-1-properties}. These parameters we are used for
numerical simulation of liquid pool profiles during vacuum arc remelting
process \cite{ken-JET}.

\begin{table}[h]
\caption{\label{table-vt3-1-properties} Properties\\ of the VT3-1 alloy.}
\begin{indented}
\item[] \begin{tabular}{@{}ll}
\br
Parameter & Value \\
\mr
$C_{s}$ & 600 J/kg K  \\
$C_{l}$ & 1200 J/kg K  \\
$\kappa_{s}$ & 10 W/m K  \\
$\kappa_{l}$ & 35 W/m K  \\
$\rho$ & 4500 kg/$m^{3}$  \\
$L$ & $3.55 \times 10^{5}$ J/kg  \\
$T_{s}$ & 1550 $^{o}C$  \\
$T_{l}$ & 1620 $^{o}C$  \\
$T_{m}$ & 1668 $^{o}C$ \\
\br
\end{tabular}
\end{indented}
\end{table}

\begin{figure}[h]
\begin{center}
\includegraphics[width=10.0cm]{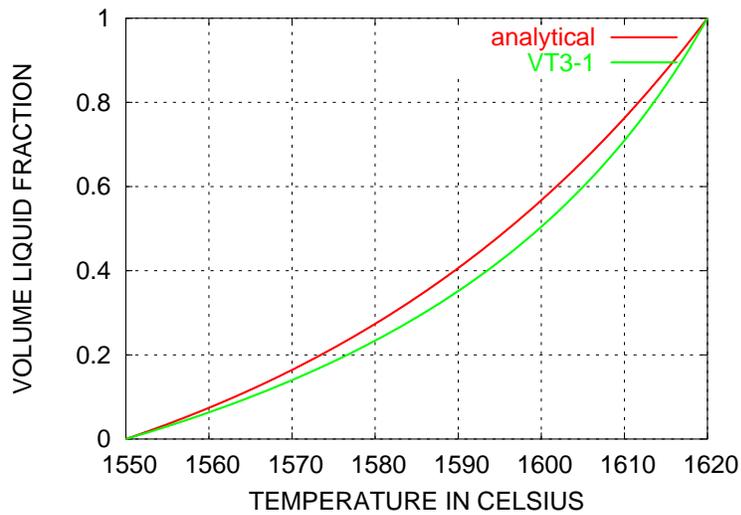}
\end{center}
\caption{\label{g_simple_alloy.eps} Liquid fraction versus temperature
for noneutectic alloy. On the Figure: {\it analytical} -- for $\lambda(T)$,
{\it VT3-1} for $\lambda_{t}(T)$.}
\end{figure}

\noindent A solution of the Eq.~(\ref{g0}) with $\lambda_{0} = 0$ is
$a_{sl} = 2.26891 \times 10^{-7} \; m^{2}/s$. Figure~\ref{g_simple_alloy.eps} shows the
temperature dependence of the
liquid fraction. Additionaly Figure~\ref{g_simple_alloy.eps} shows the function

\begin{equation}
\lambda_{t}(T) = 1 - \frac{T_{m} - T_{S}}{T_{l} - T_{s}} \cdot \frac{T_{l} - T}{T_{m} - T},
\end{equation}
which we used for VT3-1 alloy \cite{ken-JET}.
The difference between $\lambda_{T}$ and $\lambda_{t}(T)$
is small, then we have nearly realistic problem.
If $g(T)$ approximated with a power function \cite{vs}

$$
\lambda_{n} = \lb \frac{T - T_{s}}{T_{l} - T_{s}} \rb^{n},
$$
then we get $n \approx 1.5$. To test very simple numerical apparent capacity-based method
we carried out simulations with following parameters:
$a_{s} = 3.7037 \times 10^{-6} \; m^{2}/s$,
$a_{l} = 6.48148 \times 10^{-6} \; m^{2}/s$,
$T_{out} = 800 \; ^{o}C$, $T_{init} = 1650 \; ^{o}C$,
$H_{out} = 2.16\times 10^{9} \; J/m^{3}$,
$H_{init} = 10.5057 \times 10^{9} \; J/m^{3}$,
$H_{s} = 4.185 \times 10^{9} \; J/m^{3}$,
$H_{l} = 10.3455 \times 10^{9} \; J/m^{3}$.

Solutions of Eqs.~(\ref{cond-s-real})-(\ref{cond-l-real}) are $k_{s} = 0.00134109 \; m/s^{1/2}$,
$k_{l} = 0.00206009 \; m/s^{1/2}$.

\begin{figure}[h]
\begin{center}
\includegraphics[width=10.0cm]{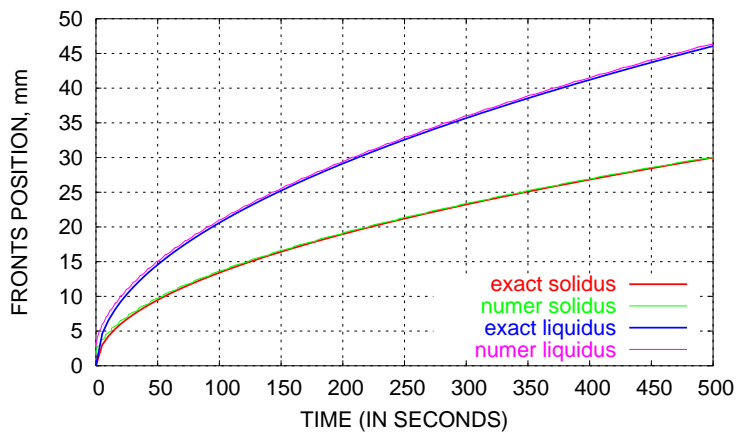}
\end{center}
\caption{\label{simple_alloy_x-t.eps}
Comparison of the apparent capacity-based numerical method with the analytical solution
as applied to the solidification of VT3-1 titanium alloy. Figure: solidus and liquidus
position versus time.}
\end{figure}

The numerical model parameters are chosen as:
length of domain $d = 0.5\;m$, nodes number $N=500$, time step $\tau=0.1\;s$.
A numerical model can provide excellent agreement with obtained
analytical solution. The results obtained are in a Figure~\ref{simple_alloy_x-t.eps}:
movement of both the solidus and the liquidus front
(\ref{Xs-Xl-tau}). We used linear interpolation between $T_{i}$ and $T_{i+1}$
for estimating position of the fronts, whereas $T_{s,l} \in [T_{i}, T_{i+1}]$.

\begin{figure}[h]
\begin{center}
\includegraphics[width=10.0cm]{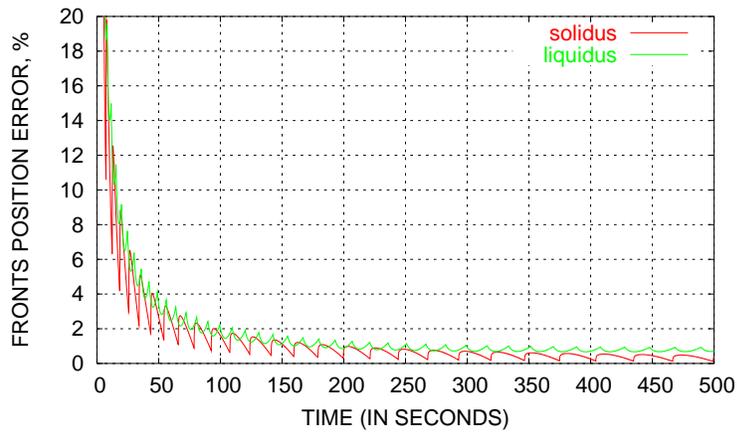}
\end{center}
\caption{\label{pos_liq_sol_err.eps}
Solidus/liquidus front position errors versus time.}
\end{figure}

\begin{figure}[h]
\begin{center}
\includegraphics[width=10.0cm]{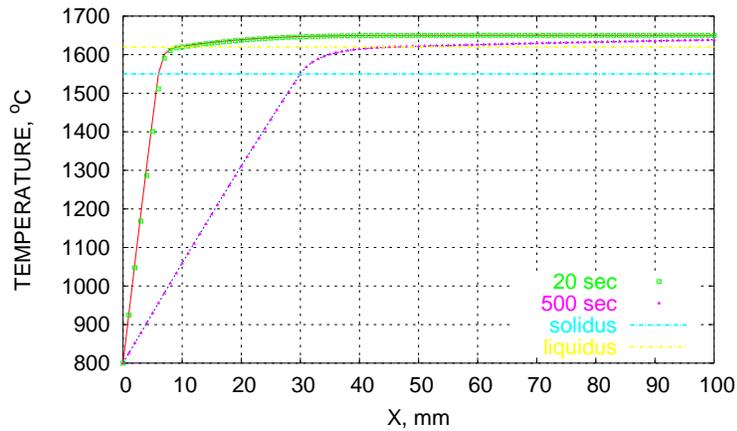}
\end{center}
\caption{\label{temp_profiles.eps}
Temperature curves at 20 and 500 seconds.}
\end{figure}

\begin{figure}[h]
\begin{center}
\includegraphics[width=10.0cm]{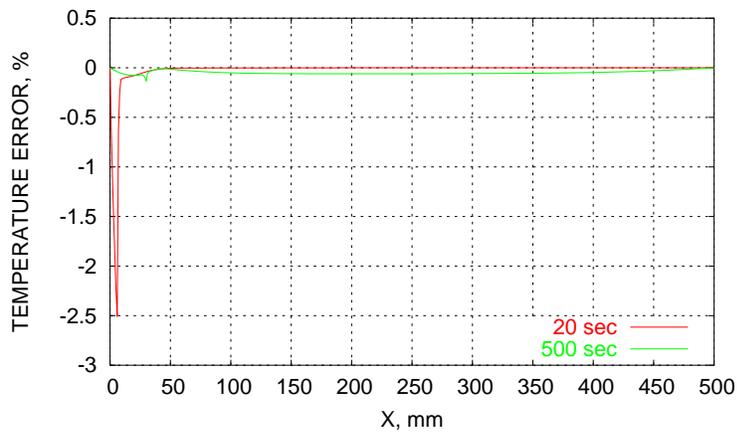}
\end{center}
\caption{\label{temp_prof_err.eps}
Temperature curves errors at 20 and 500 seconds.}
\end{figure}

The errors in the positions of solidus/liquidus
$$
\varepsilon_{x}(t) = \frac{X_{num}(t) - X_{exact}(t)}{X_{exact}(t)}\cdot 100 \%
$$
are presented in the Figure~\ref{temp_prof_err.eps}.
Moreover Figure~\ref{temp_profiles.eps} shows the temperature profiles after 20 and 500
seconds under the same numerical conditions.

Figure~\ref{temp_prof_err.eps} shows temperaure profiles errors defined by
$$
\varepsilon_{T}(x) = \frac{ T_{num}(x) - T_{exact}(x) }{ T_{exact}(x) } \cdot 100 \%.
$$

We would like to underline that the purpose of this work is to
achieve the {\em exact} analytical solution on alloy solidification.
Due to this, advantages and disatvantages different numerical algorithms can be done in
future. Due to this, we don't study the process of
solidification of $\vt$ alloy, but only use the thermo-physical properties of this
alloy to show as the model works.

\section{Conclusions.}

In this paper, analytical solution of alloy solidification problem is presented.
We developed a special method to obtain an exact analytical solution for mushy zone problem.
The main requirement of the method is thermal diffusivity to be constant in the mushy zone.
Due to such condition ordinary differential equation for
liquid fraction function is achieved. Thus the present method can be examined as ''a model'' way
to get analytical solution of some unrealistic problems.

An example of solutions is given -- the noneutectic titanium-based alloy solidification.
We provide the comparison of the simple numerical
simulation results with obtained exact solutions. We show that very our numerical apparent
capacity-based scheme provides a good agreement with exact solutions for solidus/liquidus
position and for temperatures profiles in different moments of solidification time.

Once again we would like to underline that the main goal of this paper to provide the benchmark
for binary alloy solidification problem, but not in the analysis of used numerical scheme.

If {\em predefined} $\lambda(T)$ function is to be examined, we can use another suggestions.
For example, we can require to heat conductivity (from experiment, e.g.) to be proportional
to the apparent capacity,
i.e.
$$
\kappa(T) = a_{sl} \rho \lb C_{s} + (C_{l} - C_{s})\lambda(T) +
\left[(C_{l}-C_{s})T + L\right] \frac{d\lambda(T)}{dT} \rb.
$$
Or, for the second example, we require to apparent capacity (from experiment) to be proportional to mushy heat
conductivity, i.e.
$$
\frac{dH(T)}{dT} = \frac{\kappa_{s} + (\kappa_{l}-\kappa_{s})\lambda(T)}{a_{sl}}.
$$
Moreover, we may use the B\"{a}cklund's transformation \cite{rogers-ames} to make mushy heat
equation linearisation. In this case we get {\em nonlinear} condition
$$
\frac{H^{2}(T) \lambda(T)}{\frac{dH(T)}{dT}} = const.
$$
These linearization methods will provide us with some additional analytical solutions
of alloy solidification problem.

\ack{The author thanks to Prof.~V.~Alexiades from ORNL for very usefull discussion.}


\end{document}